# Anomalous Hall Resistivity of TbBaCo$_2$O$_{5.53}$ with Non-trivial Magnetic Structure

Toshiaki Fujita, Minoru Soda and Masatoshi Sato*

*Department of Physics, Nagoya University, Furo-cho, Chikusa-ku, Nagoya 464-8602*

**Abstract**

To investigate the Hall resistivity $\rho_H$ of systems with non-trivial spin structures, transport and magnetic properties of the oxygen deficient perovskite system TbBaCo$_2$O$_{5.53}$ with non-trivial magnetic structures have been adopted. As has been already known, the system exhibits, with decreasing temperature $T$, a transition to a ferromagnetic (FM) phase with non-collinear magnetic structure at $T_C \sim 280$ K in zero magnetic field and with further decreasing $T$, the spontaneous magnetization begins to decrease at $T_1 \sim 260$ K and becomes zero at $T_N \sim 245$ K, where an antiferromagnetic (AFM) phase with non-collinear magnetic structure is realized. The metamagnetic behavior is induced by the relatively small magnetic field $H$ (< 5 T) in the $T$ region between ~200 K and $T_1$. In this $T$ region, unusual $H$ dependence of $\rho_H$ has been observed. By analyzing the observed $\rho_H$ and magnetization $M$, the $H$ dependence of the anomalous Hall coefficients $R_s$ has been obtained at various fixed temperatures. Based on the magnetic structure reported previously by our group, the possible relevance of the non-trivial magnetic structure to the observed unusual behavior of $\rho_H$ is pointed out.

## 1. Introduction

The anomalous Hall effect is one of important physical issues, which remain unsolved up to now.[1,2] As for the origin(s) of this phenomenon, there are various kinds of theoretical studies, which consider both intrinsic and extrinsic mechanisms. Karplus and Luttinger[3] proposed the existence of two intrinsic contributions to the effect, one of which is now identified to be due to the Berry phase mechanism.[4-6] As the extrinsic origins of the anomalous Hall effect, the skew-scattering[7] and side-jump mechanisms[8] are well known. In all these mechanisms, the Hall resistivity $\rho_H$ depends on the spins of the conduction electrons through the spin-orbit coupling.

The spin chirality model recently proposed by Nagaosa's group to describe the $\rho_H$ behavior observed for Nd$_2$Mo$_2$O$_7$ seemed to be a special version of the Berry phase mechanism, which can be applied to systems with the conduction electrons strongly coupled to the ferromagnetically ordered localized spins. It says that the ordered spin chirality is basically proportional to $\rho_H$, where the chirality $\chi$ is locally defined as $\chi \equiv S_1 \cdot S_2 \times S_3$ for three spins $S_1$, $S_2$ and $S_3$.[9] On the applicability of this mechanism to Nd$_2$Mo$_2$O$_7$, the present authors' group has pointed out, based on results of detailed neutron diffraction studies carried out as a function of the applied magnetic field, that the chirality mechanism cannot be a primary origin of the observed unusual behavior of $\rho_H$.[10-12] It indicates that Nd$_2$Mo$_2$O$_7$ cannot be considered to be the unique example and/or prototype of pure metals and pure compounds which exhibit the chirality-induced unusual anomalous Hall resistivity.[13] It also present a renewed problem what determines the anomalous Hall resistivity of magnetic systems.

However, because $\rho_H$ of spin-glass systems such as Fe$_{1-x}$Al$_x$, Y$_{2-x}$Bi$_x$Ir$_2$O$_7$ and SrFe$_{1-x}$Co$_x$O$_{3-\delta}$ exhibit, as we reported previously, unusual behavior, suggesting the existence of possible roles of the uniform chirality,[14-16] it seems to be still important to search for systems which present the chirality-induced anomalous Hall resistivity.

In the present work, to accumulate information on the relationship between unusual behavior of $\rho_H$ and the non-trivial spin structures, we have adopted TbBaCo$_2$O$_x$ ($x \sim 5.5$) as one of candidates exhibiting unusual behavior of $\rho_H$. The system has the alternating stacks of CoO$_6$ octahedra and CoO$_5$ square pyramids along the *b*-axis as shown in Fig 1 (a).[17-24] With decreasing temperature $T$, TbBaCo$_2$O$_{5.50}$ exhibits, in zero magnetic field, a transition to a ferromagnetic (FM) phase with non-collinear magnetic structure at $T_C \sim 280$ K. With further decreasing $T$, the spontaneous magnetization begins to decrease at $T_1 \sim 260$ K and becomes zero at $T_N \sim 250$ K, where an antiferromagnetic (AFM) phase with non-collinear magnetic structure is realized. The metamagnetic behavior is induced in relatively small external magnetic field $H$ (< 5 T) in the $T$ region between ~200 K and $T_1$. The detailed magnetic structure at $H=0$ has been studied by single crystal neutron diffraction at $T = 250$ K (in the AFM phase) and 270 K (in the FM phase): The Co$^{3+}$ ions in the CoO$_6$ octahedra are in the low spin state (spin $S = 0$), and those in the CoO$_5$ pyramids are possibly in the intermediate spin state ($S = 1$) at both temperatures.[17,18] As shown in Fig 1 (b),[17,18] both AFM and FM phases have the non-collinear magnetic structures. The Co moments in the CoO$_5$ pyramids are within the *ab*-plane and canted by the angle ~ 33.5° and ~ 45° from the *a*-axis (or *b*-axis) at $T = 250$ K (in the AFM phase) and 270 K (in the FM phase), respectively. (On these results, several groups have reported the magnetic structure and the spin state different from those reported by the authors' group. However, only the non-collinear magnetic structure reported by the present authors' group[17] can consistently explain the existing data of the neutron diffraction and other kinds of measurements, as described in ref. 18 in detail.)

If TbBaCo$_2$O$_{5.5}$ is subjected to the magnetic field along a direction out of the *ab*-plane, the local spin chirality $\chi$ is expected to be induced in the AFM phase for three neighboring spins, for example, for those connected by the dashed lines in Fig. 1 (b). According to the chirality model, the anomalous Hall resistivity proportional to $\chi$ or the fictitious flux $\Phi$ ($\propto \chi$) appears under certain conditions. In the FM phase, the local chirality is zero for any direction of the magnetic field. Keeping these facts in mind, magnetic and transport properties of polycrystalline samples of TbBaCo$_2$O$_{5+\delta}$ ($\delta$=0.50 and 0.53) have been studied in detail, where the metamagnetic behavior has been found in relatively small external magnetic field $H$ (<

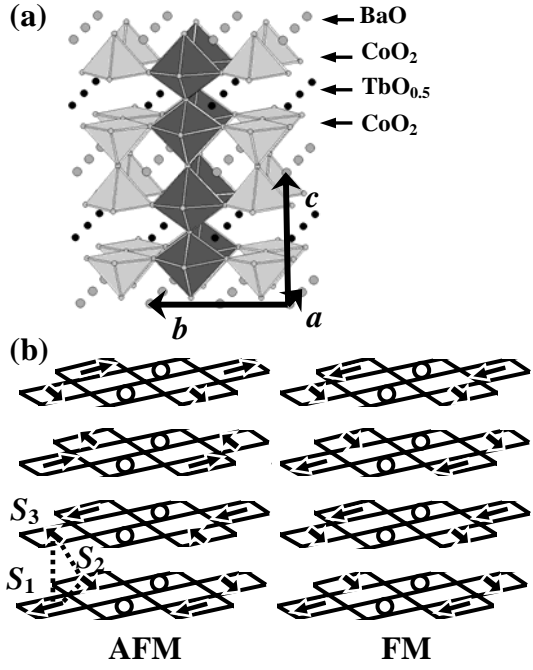

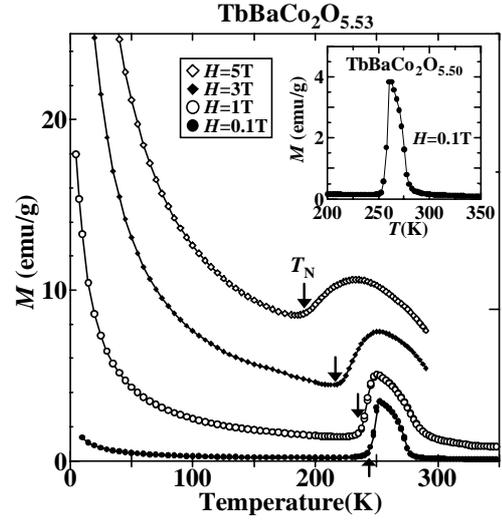

Fig. 1 (a) Oxygen deficient perovskite structure of $TbBaCo_2O_{5.5}$ is shown schematically. (b) Magnetic structures of $TbBaCo_2O_{5.5}$ reported by the present authors' group for the antiferromagnetic (AFM: left) and ferromagnetic (FM; right) phases at $H = 0$. The magnetic moments are shown by the arrows at the Co sites in the $CoO_5$ pyramids shown in (a). The open circles indicate the nonmagnetic low spin states which are in the $CoO_6$ octahedra. Three neighboring spins $S_1$, $S_2$ and $S_3$ are connected by the dashed lines. These spins induce the local spin chirality $\chi$ defined as $\chi \equiv S_1 \cdot S_2 \times S_3$.

5 T) in the $T$ region between ~200 K and $T_1$. In this $T$ region, unusual $H$ dependence of $\rho_H$ has been observed. We focus on the behavior of $\rho_H$ and show the $T$- and $H$- dependences of $\rho_H$ of $TbBaCo_2O_{5.53}$. Based on the experimental results, the relationship between the non-collinear spin structure and the unusual behavior of the anomalous Hall resistivity is discussed. We also present results of the studies on the anomalous Hall resistivities of $SrFe_{1-x}Co_xO_{3-\delta}$ ($x\sim0.5$), which has the re-entrant spin glass phases.[16]

## 2. Experiments

Polycrystalline samples of $TbBaCo_2O_{6-\delta}$ were synthesized by heating at 1150 °C for 12 h in air or flowing oxygen. The cooling rate was 100°C/h. In powder X-ray diffraction patterns taken with Fe$K\alpha$ radiation, no impurity phases were detected in these samples. They have the orthorhombic symmetry (space group P$mmm$). Their unit cells are described by ~$a_p \times 2a_p \times 2a_p$, where $a_p$ is the lattice parameter of the cubic perovskite cell, which are reported by several authors.[19-24] The oxygen contents in the samples were determined by the thermo gravimetric analysis. The oxygen contents of the sample synthesized in air is $5.50 \pm 0.02$, while those synthesized under flowing oxygen is $5.53 \pm 0.02$.

Magnetizations $M$ were measured using a Quantum Design SQUID magnetometer under a external magnetic field $H$ up to 5.5 T. Electrical resistivities $\rho$ and Hall resistivities $\rho_H$ were measured by the standard four-terminal method using a Physical Property Measurements System (PPMS, Quantum Design). In the measurements of the magnetoresistance $\Delta\rho$, $H$ was applied perpendicular to the current direction. The Hall resistivities $\rho_H$ were measured by rotating the thin plate-like sample, as usually done to remove the longitudinal component of the resistivity. In several cases, we have also measured the magnetization $M$ by rotating the sample as in the measurements of $\rho_H$.

In the studies of the $T$ dependences of $\Delta\rho$, $\rho_H$ and $M$, the samples were cooled from the temperature higher than $T_C$ to the lowest temperature (~10 K) first in zero magnetic field (zero field cooling or ZFC) and then the magnetic field $H$ was applied. At this $H$, $\rho$, $M$ or $\rho_H$ was measured with increasing $T$ to the highest temperatures of each $T$ scan. Then, for the measurements at different values of $H$, $T$ was raised above $T_C$, the field $H$ was decreased to zero by the oscillating mode and then, repeated the same procedures.

In the studies of the $H$ dependences of $\Delta\rho$, $\rho$, $\rho_H$ and $M$, the samples were cooled in zero field down to the measuring temperature $T$ from a temperature above $T_C$, and the measurements were carried out at this fixed $T$ by increasing the external fields stepwise. Then, the field $H$ was decreased to zero by the oscillating mode and then, the sample temperature was raised to a value above $T_C$ and the same procedures were repeated to take data at various fixed temperatures.

## 3. Experimental Results and Discussion

Figure 2 shows the $T$-dependence of the magnetization $M$ of a sample of $TbBaCo_2O_{5.53}$ taken under various fixed magnetic fields $H$. The inset of Fig. 2 shows the $T$-dependence of the magnetization $M$ of a sample of $TbBaCo_2O_{5.50}$ under $H = 0.1$ T. Both samples exhibit transitions with decreasing temperature $T$, to a ferromagnetic (FM) and then to an antiferromagnetic (AFM) phases. The Curie temperature $T_C$ of $TbBaCo_2O_{5.53}$ roughly estimated by using the Arrott plot ($M^2$-$H/M$ plot) is ~280 K. The $T_1$ values defined as the temperatures which the spontaneous magnetization begins to decrease with decreasing $T$, are ~ 260 K and 250 K at $H = 0.1$ T for the samples with $\delta = 0.50$ and 0.53, respectively. $T_N$ of $TbBaCo_2O_{5.53}$ decreases with increasing $H$ as shown by the arrows in Fig. 2. Magnetic structures of $TbBaCo_2O_{5.50}$ in both the AFM and FM phases at $H = 0$ are reported by the authors' group in refs. 17 and 18. We have also measured $M$ of a sample of TbBaCo$_2$O$_{5.53}$ by rotating the sample as in the measurements of ρ$_H$, and confirmed that it does not depend on the sample orientations before and after the rotation.

Fig. 2 Temperature dependences of the magnetization $M$ of $TbBaCo_2O_{5.53}$ measured with the various magnetic fields $H$. The data were taken under the zero-field-cooling condition. The arrows indicate $T_N$, defined as the characteristic temperatures, where the spontaneous magnetization becomes zero with decreasing $T$. The inset show the magnetization $M$ of $TbBaCo_2O_{5.50}$ against $T$ measured with $H = 0.1$ T under the zero-field-cooling condition.

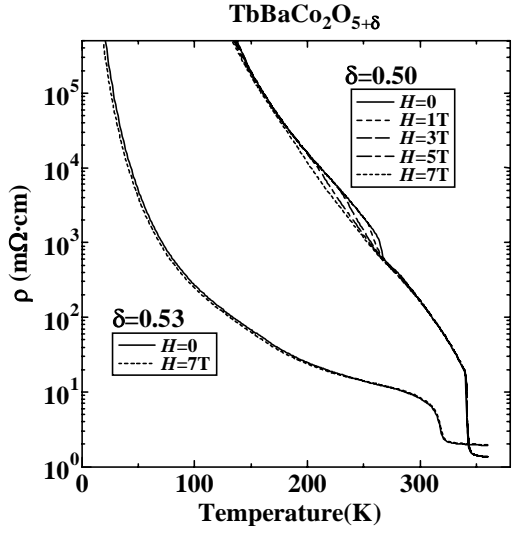

Fig. 3 Temperature dependence of the electrical resistivities ρ of TbBaCo$_2$O$_{5+\delta}$ (δ=0.50 and 0.53) measured with various magnetic fields $H$.

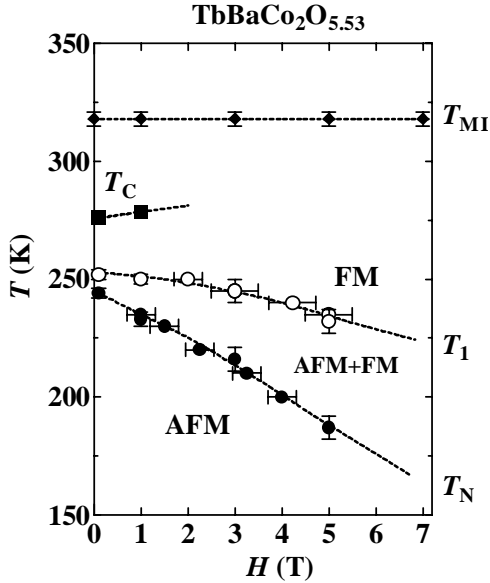

Fig. 4 Schematic $T$-$H$ phase diagram of TbBaCo$_2$O$_{5.53}$. $T_{MI}$ and $T_C$ represent the temperatures of the metal-insulator and ferromagnetic transitions, respectively. $T_N$ and $T_1$ are defined as the characteristic temperatures, where the spontaneous magnetization begins to decrease and becomes zero with decreasing $T$, respectively. The dotted lines are the guide for the eye.

Figure 3 shows the $T$-dependence of ρ of the samples of TbBaCo$_2$O$_{5+\delta}$ with δ=0.50 and 0.53 under various magnetic fields. Both samples with δ = 0.50 and 0.53 exhibit metal-insulator transitions at $T_{MI}$ ~ 340 K and 320 K, respectively. $T_{MI}$ decreases with increasing deviation δ from 0.50. This δ dependence has also been observed for other rare earth species R of RBaCo$_2$O$_{5+\delta}$.[24] $T_{MI}$ is nearly independent of $H$. At $H = 0$, ρ of TbBaCo$_2$O$_{5.50}$ exhibits the steep increase at ~$T_1$ with decreasing $T$. It can be understood to be due to the change of the double-exchange-interaction effect caused by the decrease of the ferromagnetic order parameter. This increase of ρ is significantly suppressed by the external field, because the decrease of the ferromagnetic order parameter is suppressed by the field.

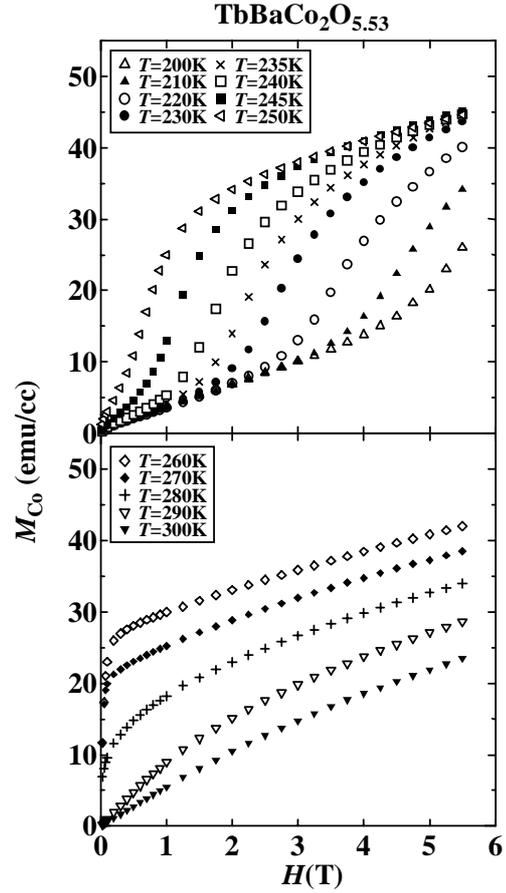

Fig. 5 Magnetizations of the Co moments $M_{Co}$ of TbBaCo$_2$O$_{5.53}$ are plotted against $H$ at various fixed temperatures. The data are taken under the ZFC condition in the range 200 K ≤ $T$ ≤ 300 K. The $M_{Co}$ values were estimated by subtracting the magnetization of TbCoO$_3$.

The resistivity ρ of TbBaCo$_2$O$_{5.53}$ does not exhibit an appreciable anomaly in the transition region. Furthermore, Δρ of TbBaCo$_2$O$_{5.53}$ in the transition region is much smaller than that of TbBaCo$_2$O$_{5.50}$, probably because the resistivity is not mainly determined by spin scattering but by the effect of randomness of the oxygen occupation. The absolute values of ρ of TbBaCo$_2$O$_{5.53}$ is smaller than those of TbBaCo$_2$O$_{5.50}$ below $T_{MI}$, possibly due to the effect of the hole doping into Co site.

In Fig. 4, $T_{MI}$, $T_1$, $T_C$ and $T_N$ values of TbBaCo$_2$O$_{5.53}$ are shown schematically against $H$. These values were obtained by measuring $M$ and ρ. The (AFM + FM) phase, where the AFM phase changes to the FM one with increasing $H$, is realized between $T_N$-$H$ and $T_1$-$H$ curves.

Figure 5 shows the $H$-dependence of the magnetization of the Co moments, $M_{Co}$ of TbBaCo$_2$O$_{5.53}$ at various fixed temperatures $T$. The values of $M_{Co}$ are estimated by subtracting $M$ of perovskite TbCoO$_3$. (Co$^{3+}$ ions of TbCoO$_3$ are in the low spin ground state and do not contribute to $M$.) In the transition region [(AFM + FM) region in Fig. 4], where the AFM phase changes to the FM one, we have observed the steep increase of $M_{Co}$ with increasing $H$ (see the upper panel).[22,24] We have not observed significant hysteretic behavior of $M_{Co}$ with varying $H$ up and down.

In Fig. 6, the $H$-dependence of Δρ of the samples TbBaCo$_2$O$_{5+\delta}$ with δ=0.50 and 0.53 are shown at several fixed temperatures. The Δρ values are negative for both samples. The magnitudes of Δρ exhibit steep increase with $H$ in the transition region, where the AFM phase changes to the FM one.[23,24]

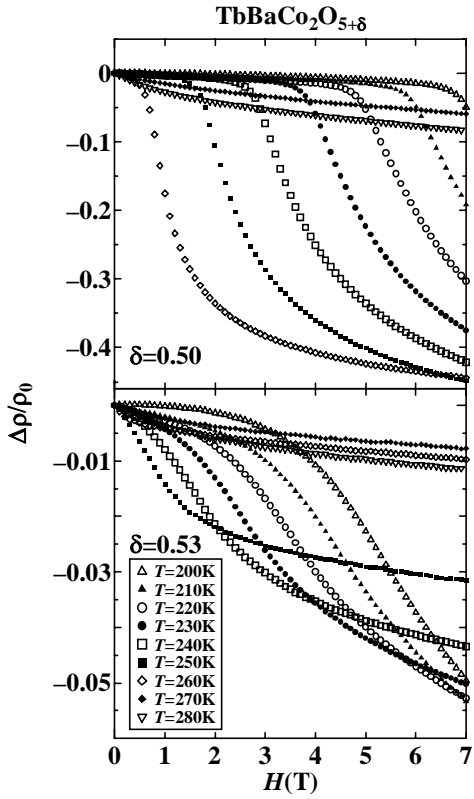

Fig. 6 Magnetoresistances $\Delta\rho/\rho_0$ of TbBaCo$_2$O$_{5+\delta}$ ($\delta$=0.50 and 0.53) are plotted against $H$ at various fixed temperatures, where $\rho_0$ is the electrical resistivities $\rho$ at $H = 0$. The data were taken under the ZFC condition.

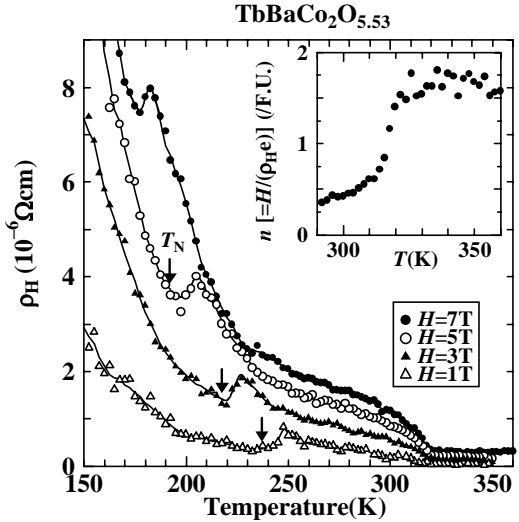

Fig. 7 Temperature dependences of Hall resistivity $\rho_H$ of TbBaCo$_2$O$_{5.53}$ under various external magnetic fields are shown. The arrows indicate the temperatures $T_N$ shown in Fig. 4. The inset shows the carrier number $n$ simply estimated using the equation $n = H/(\rho_H ec)$ at $H = 7$ T in the vicinity of $T_{MI}$.

Absolute values of $\Delta\rho/\rho_0$ of TbBaCo$_2$O$_{5.53}$ are about one order smaller than those of TbBaCo$_2$O$_{5.50}$, where $\rho_0$ is the electrical resistivity at $H$=0.

Figure 7 shows the $T$-dependence of $\rho_H$ of TbBaCo$_2$O$_{5.53}$ measured with various fixed $H$. The sign of $\rho_H$ is positive in the whole temperature region up to $T = 370$ K. The inset of Fig. 7 shows the carrier number $n$ simply estimated by the relation $n = H/(\rho_H ec)$ at $H = 7$ T, where $e$ is the electronic charge. In the figure, the values of $T_N$ are indicated by the arrows. The remarkable feature of the $\rho_H$-$T$ curves shown in the figure is the existence of the sharp increase of $\rho_H$, which begins at $T \sim T_N$ with increasing $T$. To investigate the origin of this anomalous feature, we first try to separate the observed $\rho_H$ into two components, the ordinary and anomalous Hall resistivities, measuring the $H$- and $T$-dependences of $\rho_H$.

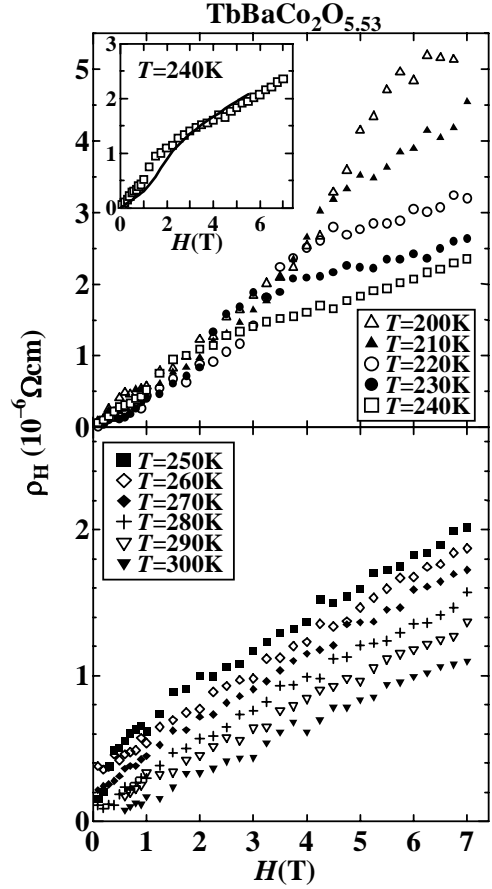

Fig. 8 $H$-dependences of the Hall resistivity $\rho_H$ of TbBaCo$_2$O$_{5.53}$ are shown at various fixed temperatures. The data are taken under the ZFC condition. The inset in the upper panel shows the $H$-dependence of $\rho_H$ of TbBaCo$_2$O$_{5.53}$ at $T = 240$ K, where the solid line is the result of the fitting of the equation $\rho_H = R_0 H + 4\pi R_s M_{Co}$ to the observed data, where $R_0$ and $R_s$ are $H$-independent.

Figure 8 shows the $H$-dependence of $\rho_H$ of the sample of TbBaCo$_2$O$_{5.53}$ at various fixed temperatures $T$. Below $T = 250$ K, we have observed the steep increase of $\rho_H$ with increasing $H$ in the transition region of $H$, where the AFM phase changes to the FM one. In the FM phase ($T = 260 \sim 270$ K), the $\rho_H$-$H$ curve seems to be described by the well-known empirical equation $\rho_H = R_0 H + 4\pi R_s M$, where $R_0$ and $R_s$ are the ordinary and anomalous Hall coefficients, respectively, and these are considered to be independent of $H$. (Strictly speaking, the equation can be used for thin plate-like samples in which the demagnetization field is well approximated by $-4\pi M$.)

The inset of the upper panel of Fig. 8 shows the $H$-dependence of $\rho_H$ at $T = 240$ K together with the fitted curve by the equation $\rho_H = R_0 H + 4\pi R_s M_{Co}$ with $H$-independent parameters $R_0$ and $R_s$. We find that the fitting is not satisfactory. Even if we consider possible $H$ dependence of $R_s$ originating from the magnetoresistance $\Delta\rho$ through the relation $R_s \propto \rho$ and/or $\propto \rho^2$,[3,7,8] we cannot obtain a satisfactory fit. To investigate the origin of this deviation from the empirical equation, we estimate the values of $H$-dependent $R_s \equiv (\rho_H - R_0 H)/(4\pi M_{Co})$ in the following way: At first, we estimated

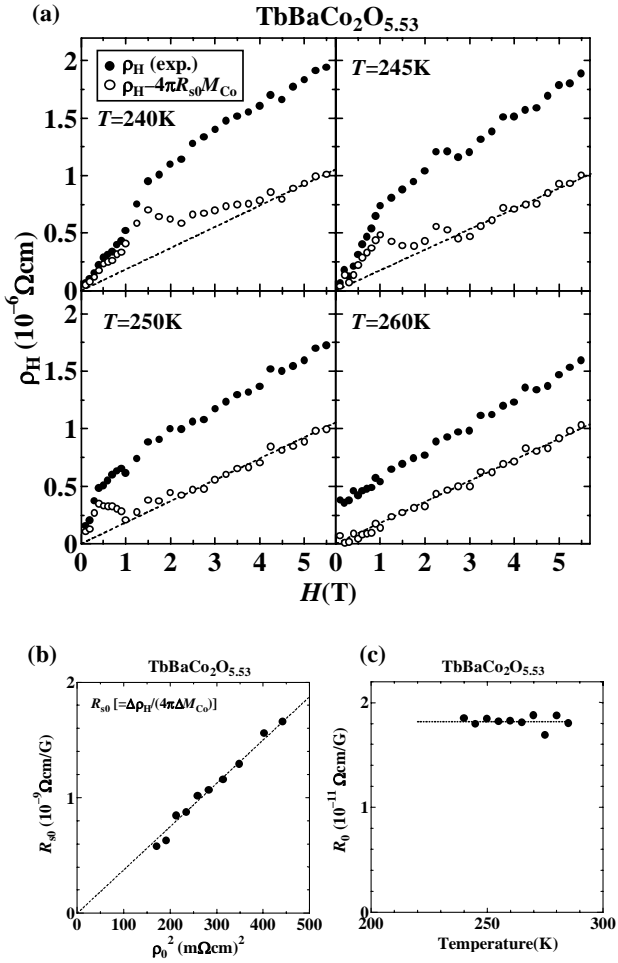

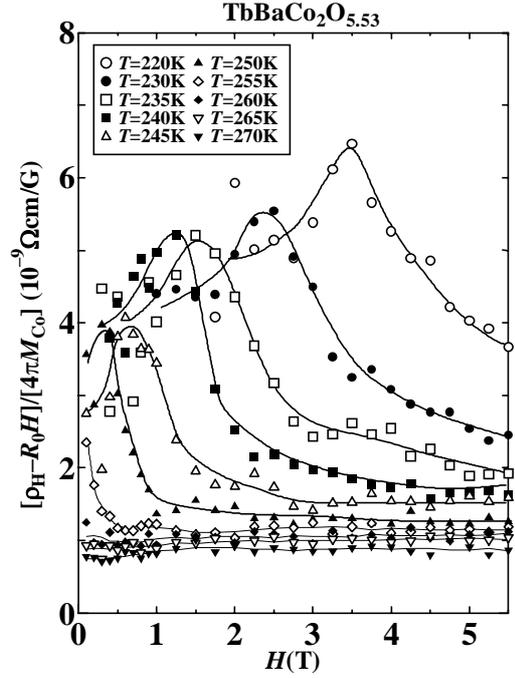

Fig. 10 $H$-dependences of the anomalous Hall coefficients $R_s = (\rho_H - R_0 H)/(4\pi M_{Co})$ of TbBaCo$_2$O$_{5.53}$ are shown at several fixed temperatures. Solid lines are the guides for the eye.

Fig. 9 (a) $H$-dependences of $\rho_H$ of TbBaCo$_2$O$_{5.53}$ at $T = 240, 245, 250$ and 260 K are shown. The filled circles indicate the raw data. The open circles represent the values of $\rho_H - 4\pi R_{s0} M_{Co}$, where $R_{s0}$ is defined as $\Delta \rho_H / (4\pi \Delta M_{Co})$, using the extrapolations $\Delta \rho_H$ and $\Delta M_{Co}$ from the high field part of the $\rho_H$-$H$ and $M_{Co}$-$H$ curves, respectively, to $H$=0. The broken lines are the results of the fitting to the data in the high field region by using the equation $\rho_H = R_0 H$. (b) $R_{s0}$ [$\equiv \Delta \rho_H / (4\pi \Delta M_{Co})$] are plotted against $\rho_0^2$, where $\rho_0$ is the electrical resistivities at $H = 0$. These data are obtained in the temperature region 240 K $\leq T \leq$ 280 K. (c) The ordinary Hall coefficients $R_0$ are plotted against $T$.

the value of $R_{s0} \equiv \Delta \rho_H / (4\pi \Delta M_{Co})$, where $\Delta \rho_H$ and $\Delta M_{Co}$ are the extrapolations to $H$=0 from the high field part of the $\rho_H$-$H$ and $M_{Co}$-$H$ curves, respectively. Next, using this estimated value of $R_{s0}$, we calculate $\rho_H - 4\pi R_{s0} M_{Co}$, which corresponds to $R_0 H$ if $R_s$ has an $H$ independent value of $R_{s0}$. The detailed $H$- dependence of $\rho_H - 4\pi R_{s0} M_{Co}$ is shown in Fig. 9(a) by the open circles. In the figure, we find that $\rho_H - 4\pi R_{s0} M_{Co}$ is nearly proportional to $H$ in the high filed region, as shown by the broken lines. The deviation from the lines can be found only in the $H$ region of the transition from the AFM to FM phases with increasing $H$, indicating that this deviation can be considered to be an unusual or anomalous behavior characteristic of the transition region.

In Fig. 9 (b), the $R_{s0}$ values obtained above are plotted against $\rho_0^2$, where $\rho_0$ is the electrical resistivities $\rho$ at $H = 0$. The data of $R_{s0}$ are taken in the temperature region 240 K $\leq T \leq$ 280 K. The relation $R_{s0} \propto \rho_0^2$ predicted by Berger[8] and Kurplus-Luttinger[3] holds well. In Fig. 9 (c), the $R_0$ values obtained above are plotted against $T$. $R_0$ is nearly independent of $T$. These results indicate that the analyses are reasonable. The carrier number $n$ estimated by the $R_0$ values [$n=1/(R_0 ec)$] is about 0.38 per formula unit, indicating that the $R_0$ value estimated in the FM region is smoothly connected to the data above 290 K (see Fig. 7).

Now, by using the $R_0$ values in Fig 9(c), we calculate the $H$ dependent $R_s$ [$\equiv (\rho_H - R_0 H)/(4\pi M_{Co})$] for arbitrary values of $H$ at various fixed temperatures. We assume that $R_0$ is independent of $H$. The results are shown in Fig. 10. In the calculations at $T = $ 235, 230 and 220 K, the averaged value of the data shown in 9(c) was used for $R_0$. It is because at these temperatures, $R_0$ could not be determined accurately by the above analyses. If we increase $H$ from a point $H$=0 in the AFM phase of Fig. 4, the data in Fig. 10 show followings: $R_s$ increases first and becomes maximum in the transition region [(AFM + FM) region in Fig 4] from the AFM phase to the FM one. With further increasing $H$, $R_s$ decreases and becomes nearly $H$ independent at the field, where it crosses the $T_1$-$H$ curve. In the FM phase, $R_s$ becomes nearly $H$ independent in the high field region. $\rho_H$ in the FM phase exhibits the ordinary behavior described by the empirical equation $\rho_H = R_0 H + 4\pi R_s M_{Co}$, where $R_0$ and $R_s$ are independent of $H$. In the measurement of $T$-dependence of $\rho_H$ shown in Fig 7, the peaks are observed in the (AFM + FM) phase, too.

As one of the candidate mechanism of the observed unusual behavior of $R_s$ in the transition region from the AFM to the FM phases, we consider a possible role of the spin chirality ($\chi$), which is related to the non-trivial magnetic structure. As for the magnetic structure of the present system, the neutron scattering studies have reported that non-collinear structures are realized in both $T$ region of the AFM and FM phases at $H = 0$.[17,18] They are shown in Fig. 1(b). As can be easily found, they do not have the local spin chirality $\chi \equiv S_1 \cdot S_2 \times S_3$ in the zero field. However, if the magnetic field with the out-of-($ab$ plane)-component is applied, we can expect that the local chirality $\chi$ or the fictitious magnetic flux $\Phi \propto \chi$ is induced. In the FM phase, for any direction of the applied magnetic field, $\chi$ is not induced, because we cannot choose the set of three spins which do not have a parallel spin pair even in the field. We cannot expect the contribution of $\chi$ to $\rho_H$ in the FM phase. However, even for the AFM phase under the field, there is no firm reason that the spin chirality mechanism contributes to the anomalous behavior of $R_s$, because the uniform chirality $\chi_0$, which is sum of the local spin chirality, is zero. Resultingly, $\rho_H \propto \chi_0$ cannot be induced in any of the AFM and FM phases.

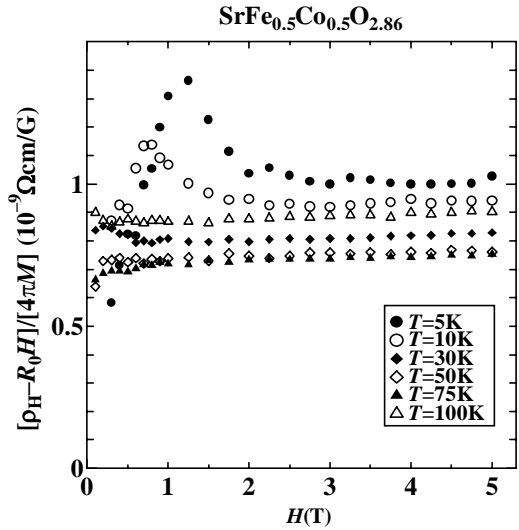

Fig. 11 *H*-dependences of the anomalous Hall coefficients $R_s=(\rho_H-R_0H)/(4\pi M)$ of SrFe$_{0.5}$Co$_{0.5}$O$_{2.86}$ are shown at several fixed temperatures.

However, for the present polycrystalline samples of TbBaCo$_2$O$_{5+\delta}$, the excess $R_s$ induced by the external field is experimentally observed in the transition region. We presume that $\chi_0$ becomes nonzero in the coexisting state of the AFM and FM phases. This idea seems to be supported by the fact that the anomalous behavior of the $R_s$ which may be related to the nonzero $\chi_0$ has been observed in spin glass systems.[14-16] In paticular, the quite similar behavior of $R_s$ has been reported for SrFe$_{1-x}$Co$_x$O$_{3-\delta}$.[16] In this system, the re-entrant spin glass behavior is very significant below $T \equiv T_g \sim 30$ K for $0.2 \leq x \leq 0.6$.[16] As can be found in ref. 16, the characteristic deviation of the $\rho_H$-$H$ curves from the relation $\rho_H = R_0H + 4\pi R_sM$, where $R_0$ and $R_s$ are independent of $H$, can be found in the re-entrant spin glass phase. Here, we calculate the $H$ dependent anomalous Hall coefficients $R_s \equiv (\rho_H-R_0H)/(4\pi M)$ using the data of SrFe$_{1-x}$Co$_x$O$_{3-\delta}$ ($x=0.5$, $\delta=0.14$) in the manner described above and show the results in Fig. 11. The behavior of the $R_s$-$H$ curve is quite similar to that of the present TbBaCo$_2$O$_{5.53}$. The field of the maximum $R_s$ increases with decreasing $T$. At $T = 50, 75$ and 100 K in the FM phase, $R_s$ is nearly independent of $H$. On this kind of unusual behavior of the Hall resistivity, Tatara and Kawamura proposed that the uniform chirality induce by the magnetic field contributes to the Hall resistivity.[25,26]

It is plausible in the present TbBaCo$_2$O$_{5.53}$ that the uniform chirality induced in the transition region contributes to the observed excess component of $R_s$.

## 4. Conclusion

The transport and magnetic studies on polycrystalline samples of the oxygen deficient perovskite TbBaCo$_2$O$_{5.53}$ have been carried out, focusing on the behavior of the Hall resistivity $\rho_H$. In the transition region where the AFM phase changes to the FM one, we have observed the deviation of the $\rho_H$-$H$ curves from the empirical relation $\rho_H = R_0H + 4\pi R_sM$ with $H$-independent $R_0$ and $R_s$. By analyzing $\rho_H$ and $M$ in this region, we have extracted unusual $H$-dependences of the anomalous Hall coefficients $R_s$. Based on the results of the previous magnetic structure analyses, we have discussed the possible role of the non-collinear spin structure in the determination of the anomalous Hall resistivities. We have pointed out that in the re-entrant spin glass phase of SrFe$_{1-x}$Co$_x$O$_{3-\delta}$,[16] similar unusual behaviors of the anomalous Hall resistivities has been observed.

Acknowledgements – The work is supported by Grants-in-Aid for Scientific Research from the Japan Society for the Promotion of Science (JSPS) and by Grants-in-Aid on priority area from the Ministry of Education, Culture, Sports, Science and Technology.

**References**
1) J.-P. Jan: Helv. Phys. Acta **25** (1952) 677.
2) J. J. Rhyne: Phys. Rev. **172** (1968) 523.
3) R. Kurplus and J. M. Luttinger: Phys. Rev. **95** (1954) 1154.
4) G. Sundaram and Q. Niu: Phys. Rev. B **59** (1999) 14915.
5) M. Onoda and N. Nagaosa: J. Phys. Soc. Jpn. **71** (2002) 19.
6) T. Jungwirth, Q. Niu and A. H. MacDonald: Phys. Rev. Lett. **88** (2002) 207208.
7) J. Smit: Physica **21** (1955) 877.
8) L. Berger: Phys. Rev. B **2** (1970) 4559.
9) K. Ohgushi, S. Murakami and N. Nagaosa: Phys. Rev. B **62** (2000) R6065.
10) Y. Yasui, Y. Kondo, M. Kanada, M. Ito, H. Harashina, M. Sato and K. Kakurai: J. Phys. Soc. Jpn. **70** (2001) 284.
11) Y. Yasui, Y. Kondo, M. Kanada, M. Ito, H. Harashina, S. Iikubo, K. Oda, T. Kageyama, K. Murata, S. Yoshii, M. Sato, H. Okumura and K. Kakurai: *Proc. 1st Int. Symp. Advanced Science Research*, *Advances in Neutron Scattering Research*, J. Phys. Soc. Jpn. **70** (2001) Suppl. A., p.100.
12) Y. Yasui, S. Iikubo, H. Harashina, T. Kageyama, M. Ito, M. Sato and K. Kakurai: J. Phys. Soc. Jpn. **72** (2003) 865.
13) Y. Yasui, T. Kageyama, T. Moyoshi, M. Soda, M. Sato and K. Kakurai: J. Phys. Soc. Jpn. **75** (2006) No. 8.
14) T. Kageyama, N. Aito, S. Iikubo and M. Sato: J. Phys. Soc. Jpn. **72** (2003) 1491.
15) N. Aito, M. Soda, Y. Kobayashi and M. Sato: J. Phys. Soc. Jpn. **72** (2003) 1226.
16) T. Ido, Y. Yasui and M. Sato: J. Phys. Soc. Jpn. **72** (2003) 357.
17) M. Soda, Y. Yasui, T. Fujita, T. Miyashita, M. Sato and K. Kakurai: J. Phys. Soc. Jpn. **72** (2003) 1729.
18) M. Soda, Y. Yasui, Y. Kobayashi, T. Fujita, M. Sato and K. Kakurai: submitted to J. Phys. Soc. Jpn.
19) A. Maignan, C. Martin, D. Pelloquin, N. Nguyen and B. Raveau: J. Solid. State Chem. **142** (1999) 247.
20) D. Akahoshi and Y. Ueda: J. Solid. State Chem. **156** (2001) 355.
21) H. Kusuya, A. Machida, Y. Moritomo, K. Kato, E. Nishibori, M. Takata, M. Sakata and A. Nakamura: J. Phys. Soc. Jpn. **70** (2001) 3577.
22) M. Baran, V. I. Gatalskaya, R. Szymczak, S. V. Shiryaev, S. N. Barilo, K. Piotrowski, G. L. Bychkov and H Szymczak: J. Phys. Condens. Matter **15** (2003) 8853.
23) Z. X. Zhou and P. Schlottmann: Phys. Rev. B **71** (2005) 174401.
24) A. A. Taskin, A.N. Lavrov and Y. Ando: Phys. Rev. B **71** (2005) 134414.
25) G. Tatara and H. Kawamura: J. Phys. Soc. Jpn. **71** (2002) 2613.
26) H. Kawamura: Phys. Rev. Lett. **90** 047202 (2003)